# Search for Gamma-Ray Bursts and Gravitational Wave Electromagnetic Counterparts with High Energy X-ray Telescope of *Insight*-HXMT

C. Cai[1,2]★, S. L. Xiong[1]†, C. K. Li[1], C. Z. Liu[1], S. N. Zhang[1,2], X. B. Li[1], L. M. Song[1,2], B. Li[1], S. Xiao[1,2],
Q. B. Yi[1,7], Y. Zhu[1], Y. G. Zheng[1,6], W. Chen[1], Q. Luo[1,2], Y. Huang[1], X. Y. Song[1], H. S. Zhao[1], Y. Zhao[1],
Z. Zhang[1], Q. C. Bu[1], X. L. Cao[1], Z. Chang[1], L. Chen[4], T. X. Chen[1], Y. B. Chen[3], Y. Chen[1], Y. P. Chen[1],
W. W. Cui[1], Y. Y. Du[1], G. H. Gao[1,2], H. Gao[1,2], M. Y. Ge[1], Y. D. Gu[1], J. Guan[1], C. C. Guo[1,2], D. W. Han[1],
J. Huo[1], S. M. Jia[1], W. C. Jiang[1], J. Jin[1], L. D. Kong[1,2], G. Li[1], T. P. Li[1,3], W. Li[1], X. Li[1], X. F. Li[1], Z. W. Li[1],
X. H. Liang[1], J. Y. Liao[1], B. S. Liu[1], H. W. Liu[1], H. X. Liu[1], X. J. Liu[1], F. J. Lu[1], X. F. Lu[1], T. Luo[1], R. C. Ma[1,2],
X. Ma[1], B. Meng[1], Y. Nang[1,2], J. Y. Nie[1], G. Ou[1], J. L. Qu[1,2], X. Q. Ren[1,2], N. Sai[1,2], L. Sun[1], Y. Tan[1],
L. Tao[1], Y. L. Tuo[1], C. Wang[2,5], L. J. Wang[1], P. J. Wang[1,2], W. S. Wang[1], Y. S. Wang[1], X. Y. Wen[1], B. B. Wu[1],
B. Y. Wu[1,2], M. Wu[1], G. C. Xiao[1], Y. P. Xu[1], R. J. Yang[6], S. Yang[1], Y. J. Yang[1], Y. R. Yang[1], X. J. Yang[7],
Q. Q. Yin[1], Y. You[1,2], F. Zhang[1], H. M. Zhang[1], J. Zhang[1], P. Zhang[1], S. Zhang[1], W. C. Zhang[1], W. Zhang[1,2],
Y. F. Zhang[1], Y. H. Zhang[1,2], X. F. Zhao[1,2], S. J. Zheng[1], D. K. Zhou[1,2]

[1] *Key Laboratory of Particle Astrophysics, Institute of High Energy Physics, Chinese Academy of Sciences, Beijing 100049, China*
[2] *University of Chinese Academy of Sciences, Chinese Academy of Sciences, Beijing 100049, China*
[3] *Department of Astronomy, Tsinghua University, Beijing 100084, China*
[4] *Department of Astronomy, Beijing Normal University, Beijing 100088, China*
[5] *Key Laboratory of Space Astronomy and Technology, National Astronomical Observatories, Chinese Academy of Sciences, Beijing 100012, China*
[6] *College of physics Sciences & Technology, Hebei University, No. 180 Wusi Dong Road, Lian Chi District, Baoding City, Hebei Province 071002, China*
[7] *Department of Physics, Xiangtan University, Xiangtan, Hunan Province 411105, China*

**ABSTRACT**
The High Energy X-ray telescope (HE) on-board the Hard X-ray Modulation Telescope (*Insight*-HXMT) can serve as a wide Field of View (FOV) gamma-ray monitor with high time resolution ($\mu$s) and large effective area (up to thousands cm$^2$). We developed a pipeline to search for Gamma-Ray Bursts (GRBs), using the traditional signal-to-noise ratio (SNR) method for blind search and the coherent search method for targeted search. By taking into account the location and spectrum of the burst and the detector response, the targeted coherent search is more powerful to unveil weak and sub-threshold bursts, especially those in temporal coincidence with Gravitational Wave (GW) events. Based on the original method in literature, we further improved the coherent search to filter out false triggers caused by spikes in light curves, which are commonly seen in gamma-ray instruments (e.g. *Fermi*/GBM, *POLAR*). We show that our improved targeted coherent search method could eliminate almost all false triggers caused by spikes. Based on the first two years of *Insight*-HXMT/HE data, our targeted search recovered 40 GRBs, which were detected by either *Swift*/BAT or *Fermi*/GBM but too weak to be found in our blind search. With this coherent search pipeline, the GRB detection sensitivity of *Insight*-HXMT/HE is increased to about 1.5E-08 erg/cm$^2$ (200 keV–3 MeV). We also used this targeted coherent method to search *Insight*-HXMT/HE data for electromagnetic (EM) counterparts of LIGO-Virgo GW events (including O2 and O3a runs). However, we did not find any significant burst associated with GW events.

## 1 INTRODUCTION

The association between gravitational wave (GW 170817) (Abbott et al. 2017a) and short GRB (GRB 170817A) (Goldstein et al. 2017; Savchenko et al. 2017) that are produced by the merger of double neutron stars heralded the multi-messenger gravitational wave astronomy era. As the China's first X-ray astronomy satellite (Zhang 2009) launched on June 15, 2017, the *Hard X-ray Modulation Telescope* (*Insight*-HXMT) (Zhang et al. 2019) is equipped with three main instruments (Zhang et al. 2014): the High Energy X-ray telescope (HE) (Liu et al. 2019), the Medium Energy X-ray telescope (ME) (Cao et al. 2019) and the Low Energy X-ray telescope (LE) (Chen et al. 2019). During the early days of its commissioning operation, the *Insight*-HXMT monitored the location region of GW 170817 for prompt gamma-ray emission with HE, followed by a Target of Opportunity (ToO) scanning observation to cover a large region of the GW location error with all three telescopes (Li et al. 2018). As a result of these observations, HE provided a stringent constraint on the potential MeV emission of the GRB 170817A (Abbott et al. 2017b).

★ E-mail: caice@ihep.ac.cn
† E-mail: xiongsl@ihep.ac.cn

Thanks to the large effective area (up to thousands cm$^2$ (Luo et al. 2020)), large FOV (i.e. un-occulted sky by the Earth), high time resolution (microseconds) and short deadtime (Xiao et al. 2020), HE is a versatile gamma-ray monitor (0.2–3 MeV) for GRBs as well as the possible gamma–ray counterparts of Fast Radio Bursts (FRBs) (Guidorzi et al. 2020b,a). Indeed, using all three telescopes, *Insight*-HXMT discovered and measured, in a great detail, the first X-ray burst associated with FRB 200428 from the Galactic magnetar SGR J1935+2154 (Li et al. 2021).

Although there is no real-time on-board trigger or data telemetry for *Insight*-HXMT satellite, we have developed a pipeline to do ground search for GRBs. The typical time latency for telemetry data and the corresponding ground search is about 10 hours. Both blind search and targeted search are implemented in our pipeline. The blind search looks for signals with an excess signal-to-noise ratio (SNR) in HE detectors.

Although most GRBs can be found through blind search, in order to improve the capability to uncover weak bursts, we also used the targeted coherent search method in our pipeline, which was proposed by Blackburn et al. (2015). With a prior information about trigger time and/or location, which were provided by other instruments, the coherent search would estimate the expected counts for each energy channel of each detector by combining the possible burst locations, burst spectra and instrument responses, then compare the expected counts with the observed counts to construct a likelihood ratio. This method yields a more sensitive search than the signal-to-noise ratio method.

In this paper, we present the methodology and results of the ground search of HE data for both GRBs and high energy counterparts of GW events. After a brief description of the traditional search method for blind search, we focus on the coherent method for targeted search to find weak GRBs in HE data. This paper is structured as follows: The HE telescope is introduced in section 2. The blind search and targeted search of GRBs are presented in section 3 and section 4 respectively. The search result for gravitational wave electromagnetic counterpart is shown in section 5. Finally, we give discussion in section 6 and summary in section 7.

## 2 HE TELESCOPE OF *INSIGHT*-HXMT

*Insight*-HXMT/HE is composed of 18 cylindrical NaI(Tl)/CsI(Na) phoswich detectors, with a total geometrical area of about 5100 cm$^2$. The measured energy (also called deposited energy) range of CsI (NaI) is about 40−600 keV (20−250 keV) in the normal gain mode and about 200 keV−3 MeV (100 keV−1.25 MeV) in the low gain mode (Zhang et al. 2019; Liu et al. 2019; Li et al. 2019).

NaI is used as the primary detector to measure ∼20-250 keV photons incident from the field of view (FOV) defined by collimator, for observing spectra and temporal variability of X-ray sources either by pointing observations for known sources or scanning observations to unveil new sources, while CsI is originally designed to work as anti-coincident detector for NaI. However, gamma-rays with energy greater than about 200 keV and any incident angle with respect to the HE telescope, can penetrate the satellite structure and deposit energy in the CsI detector (the probability of energy deposition in NaI is very low), which means that HE CsI could serve as an all-sky gamma-ray monitor for GRBs and the gamma-ray counterpart of various sources, such as GW events and FRBs.

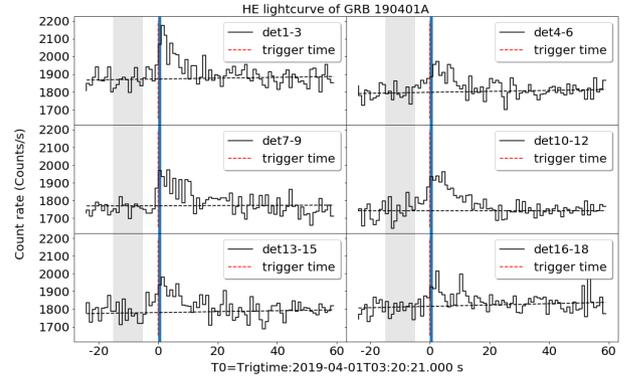

**Figure 1.** *Insight*-HXMT/HE CsI light curves of GRB 190401A, which triggered the blind search. Each panel corresponds to each group of CsI detectors. The source region (1 s) and background region are shown by the blue and gray shaded regions, respectively. Trigger time is marked with the red dotted lines. The horizontal dash lines are the estimated background.

For each NaI/CsI phoswich detector of HE, the event-by-event data (called EVT data hereafter) is recorded and we can use the pulse width information in the EVT data to distinguish between NaI events and CsI events. For this paper, we only select the events recorded by CsI detectors by screening events with pulse width greater than 75 (Liu et al. 2019). We only make use of the HE CsI data in the first two years of *Insight*-HXMT (from June 15, 2017 to June 30, 2019) in this work.

## 3 BLIND SEARCH FOR GRBS

The blind search pipeline is based on the signal-to-noise ratio (i.e. SNR) method which is applied to the light curves. To reduce statistical fluctuations, 18 CsI detectors of HE are divided into 6 groups, and we derived the total light curve for each group by summing up all 3 CsI detectors in each group. For each time interval (source region hereafter), we determine the significance of the excess over background fluctuation for each light curve for all 6 detector groups, by using the following equations:

$$N_{\text{bkg}} = N'_{\text{bkg}} \times \frac{T_{\text{total}}}{T_{\text{bkg}}}, \quad (1)$$

where $N'_{\text{bkg}}$ and $T_{\text{bkg}}$ are the total counts and duration of the pre-burst background region, respectively. $T_{\text{total}}$ and $N_{\text{bkg}}$ are the duration and estimated background counts of the source region, respectively. Then the significance can be estimated as:

$$Sig = \frac{N_{\text{net}}}{\sigma_{N_{\text{bkg}}}} = \frac{N_{\text{total}} - N_{\text{bkg}}}{\sqrt{N_{\text{bkg}}}}, \quad (2)$$

where $N_{\text{total}}$ is the total counts in the source region.

A trigger is achieved when more than 3 groups of CsI detector having significance higher than $3\sigma$, according to our settings. This blind search is implemented for the following time scales (i.e. $T_{\text{total}}$): 0.05 s, 0.1 s, 0.2 s, 0.5 s and 1 s. The source regions are defined as the five time scales mentioned above. Based on many tests on HE CsI data, the background region is chosen from $T_0$-15 s to $T_0$-5 s, which is shown in Figure 1. We fit the background with a polynomial function and found that the fitting with order 0 is more applicable to the HE CsI data. As an example of the blind search result, light curves of GRB 190401A are shown in Figure 1, which was also observed by *Fermi*/GBM.

During the first two years of *Insight*-HXMT operation, the blind search pipeline has found 180 GRBs in HE CsI data. Detailed analysis of these *Insight*-HXMT/HE GRBs will be published in the forthcoming GRB catalogue paper ( X. Y. Song et al. submitted).

## 4 TARGETED SEARCH FOR GRBS

### 4.1 GRB Samples for targeted search

In order to do targeted search in HE CsI data, we used those GRBs that were detected by *Swift*/BAT or *Fermi*/GBM but not found by HE CsI with blind search from June 15, 2017 to June 30, 2019. As one of the three instruments on-board the *Neil Gehrels Swift Observatory* (Barthelmy et al. 2005), BAT is a large (5200 cm$^2$) CZT-based coded aperture imager, covering a FoV of 1.4 sr. BAT can measure the spectrum in 15–150 keV and provide accurate location (1-4 arcminutes of error) for GRBs.

The GBM onboard the *Fermi Gamma-Ray Space Telescope* consists of twelve sodium iodide (NaI) detectors (8 keV to ∼1000 keV) and two bismuth germanate (BGO) crystal detectors (200 keV to ∼40 MeV) (Meegan et al. 2009). Unlike the coded mask imaging used by BAT, GBM reconstructs GRB location using the relative counts of all detectors compared to pre-defined location templates (Connaughton et al. 2015). The GBM location error of GRBs are usually several degrees depending on many factors, such as the brightness of the GRBs, incident angle with respect to the spacecraft and so on. Although GBM's location has much larger error than BAT, GBM's energy range (8 keV to ∼40 MeV) is much wider than BAT (15–150 keV), thus the GBM spectrum is more applicable in the energy band (0.2–3 MeV) of HE CsI gamma-ray monitoring.

The GBM flight software would give rise to a trigger when a statistically significant excess (typically $4.5\sigma$ threshold) in the time-binned data occurs in at least two NaI detectors (Burns et al. 2019). In order to recover weak bursts that can not be found by GBM flight software, a ground-based blind search is designed by GBM team to search the continuous time-tagged event (CTTE) data. This blind search covers much wider energy band and a much larger range of time scale than that of the flight software. In addition, a more sensitive targeted coherent search was developed by GBM team to unveil much weaker bursts (see Blackburn et al. (2013) for more details).

In this paper, we defined three GRB samples: GRBs detected by BAT only, detected by GBM only and detected by both BAT and GBM, which are denoted as BAT-only, GBM-only and BAT+GBM, respectively. There are 627 GRBs in these three samples before any selection. Next we only choose those GRBs for which *Insight*-HXMT/HE CsI has collected data and was unblocked by the Earth with respect to the GRB location.

For the BAT-only and BAT+GBM samples, using the precise location provided by BAT, we can select those GRBs that are certainly visible to *Insight*-HXMT, i.e., without Earth blocking. This results in a sample of 9 short bursts and 91 long GRBs, where 3 short GRBs and 24 long GRBs were detected by both BAT and GBM.

As for the GBM-only sample, because GBM location error is relatively large, we just select those GRBs for which the full 3-$\sigma$ location error region is not blocked by the Earth for *Insight*-HXMT. This results in 155 long GRBs and 33 short GRBs.

**Table 1.** Three GRB samples for targeted search in this work

| Sample | Long GRBs | Short GRBs | Total |
| --- | --- | --- | --- |
| BAT-only | 67 | 6 | 73 |
| GBM-only | 155 | 33 | 188 |
| GBM+BAT | 24 | 3 | 27 |
| total | 246 | 42 | 288 |

**Table 2.** Model Band Spectral Parameters

| Spectrum | $\alpha$ | $\beta$ | $E_{\text{peak}}$ |
| --- | --- | --- | --- |
| soft | -1.9 | -3.7 | 70 keV |
| normal | -1 | -2.3 | 230 keV |
| hard | 0 | -1.5 | 1 MeV |

Finally, we have a total of 288 GRBs for targeted search, which are summarized in Table 1. For these GRBs, we retrieved the trigger time, burst duration ($T_{90}$), location and the best fit spectral model from the *Fermi* GBM Burst Catalog (Gruber et al. 2014; von Kienlin et al. 2014; Bhat et al. 2016; von Kienlin et al. 2020) and the Third *Swift* BAT GRB Catalog (Lien et al. 2016).

### 4.2 Coherent search method

We applied different search time windows for short GRBs and long GRBs in our samples: $T_0 \pm 5$ s for short bursts and $T_0 \pm 50$ s for long bursts, centered on the trigger time ($T_0$) of GBM or BAT. This search is performed on HE CsI event data, covering a series of timescales ranging from 0.05 s to 4 s, with 8 energy channels and two phases offset.

The search first estimates the expected counts rate for each energy channel, each detector and each possible location in the sky at a given time, using three templates of GRB photon spectrum folded through detector response. Three types of Band function (Band et al. 1993) (denoted as soft Band, normal Band and hard Band) are adopted as the typical GRB spectra incident to HE, with parameters $\alpha$, $\beta$ and $E_{\text{peak}}$ shown in Table 2 (Connaughton et al. 2015).

Following the methodology of coherent search in Blackburn et al. (2013), we construct the likelihood by comparing the expected counts to the measured counts in each channel ($i$) and each detector ($k$). Finally, a likelihood ratio is defined as the ratio between the probability of measuring the observed data in the presence of a burst signal and the probability of measuring the observed data with background only. $H_0$ represents the hypothesis that there is only background, while $H_1$ represents that there is a burst signal.

For each detector $k$ and each channel $i$, assuming uncorrelated Gaussian noise,

$$P_k(d_k|H_1) = \prod_i \frac{1}{\sqrt{2\pi}\sigma_{d_i}} exp(-\frac{(\widetilde{d}_i - r_i s)^2}{2\sigma_{d_i}^2}), \qquad (3)$$

$$P_k(d_k|H_0) = \prod_i \frac{1}{\sqrt{2\pi}\sigma_{n_i}} exp(-\frac{\widetilde{d}_i^2}{2\sigma_{n_i}^2}), \qquad (4)$$

$$\widetilde{d}_i = d_i - \langle n_i \rangle, \qquad (5)$$

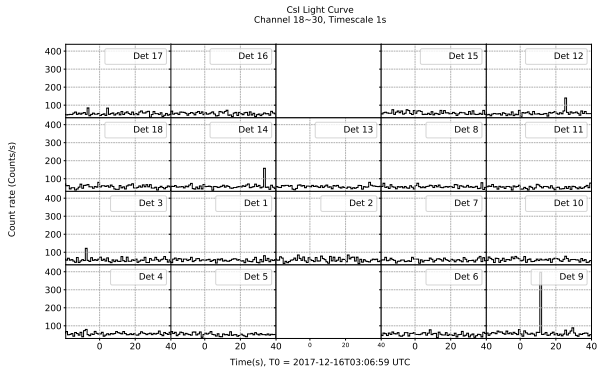
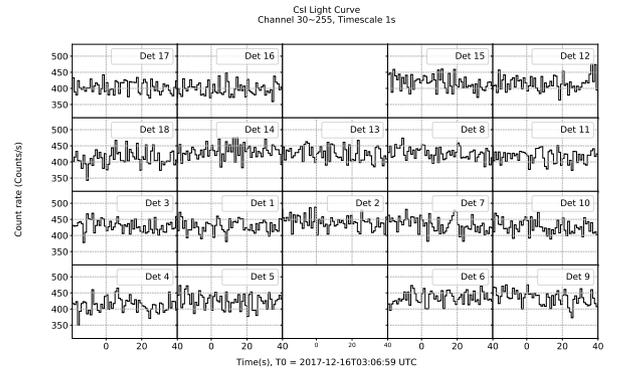

**Figure 2.** *Insight*-HXMT/HE CsI Light curves of GRB 171216A in different channels (left panel: 18~30; right panel:30~255 ), which are distributed according to the position of the detectors installed on the satellite. T0 is the trigger time of GRB 171216A, which was detected by *Swift*/BAT. There is a spike appearing near the trigger time in the lower channel of CsI detector #9.

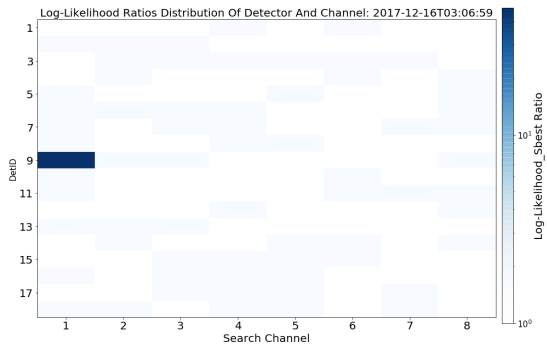

**Figure 3.** Log-Likelihood ratio distribution of different detectors and different channels for the spike event, which occurred on a single detector (CsI detector #9). This trigger can be identified by the maximum $\mathcal{L}$ value ratio of one detector to the total detectors.

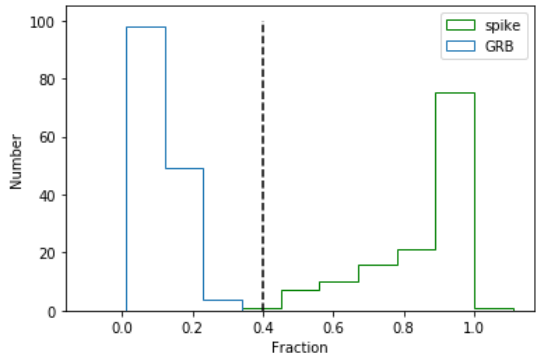

**Figure 4.** The distribution of fraction ($f_{\max}$) of the maximum log-Likelihood ratio of a single detector ($\mathcal{L}_k$) to the summed log-Likelihood ratio of all detectors ($\mathcal{L}$). The blue and green line represent the distribution for GRBs and spikes. The black dotted line represents the threshold value (0.4).

where the product is carried out over each channel, $d_i$ represents the observed counts, $n_i$ represents the estimated background, $\sigma_{n_i}$ and $\sigma_{d_i}$ represent the standard deviation of the background data and expected data (background+signal) respectively, $\widetilde{d_i}$ represents background-subtracted data, $r_i$ represents the instrument response, depending on the direction and energy spectrum of source and $s$ is the intrinsic source amplitude (if there is only noise, $s$=0).

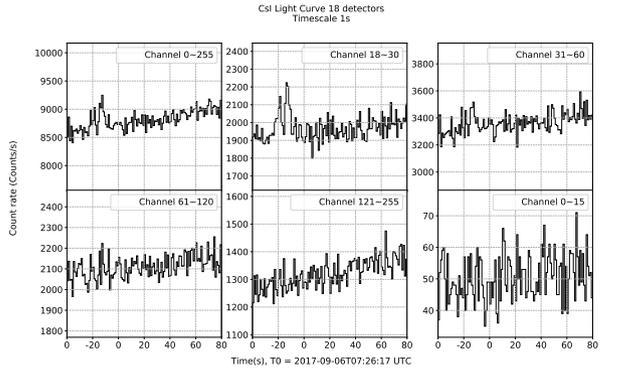

**Figure 5.** *Insight*-HXMT/HE CsI light curves of GRB 170906C in different energy channels. This burst has most signals in channels 18~30. T0 is the BAT trigger time of GRB 170906C. This GRB triggered the targeted search successfully by using our improved coherent search method of excluding spike events. This burst will not trigger if removing the low energy channels in HE CsI data.

The final log likelihood ratio is the summed likelihood ratio of each detector:

$$\mathcal{L}_k = \ln \frac{P_k(d_k|H_1)}{P_k(d_k|H_0)} = \sum_{i=1}[\ln \frac{\sigma_{n_i}}{\sigma_{d_i}} + \frac{\widetilde{d_i}^2}{2\sigma_{n_i}^2} - \frac{(\widetilde{d_i} - r_i s)^2}{2\sigma_{d_i}^2}], \quad (6)$$

$$\mathcal{L} = \sum_{k=1} \mathcal{L}_k. \quad (7)$$

According to the tests, we set trigger threshold of $\mathcal{L}$ to be 9, as proposed in Kocevski et al. (2018). For a burst source, the likelihood ratio varies with the spectrum and location of the source, as well as the search time window. We choose the highest likelihood ratio as the final result of the search.

Just like other gamma-ray instruments (Kouveliotou et al. 1992; Fishman & Austin 1977), spikes are also commonly seen in the light curves of HE and are likely caused by the interactions of high energy cosmic rays with detectors. Those spikes usually occur in a single detector with high counts rate (most of them have duration less than 200 us). The events in spikes always have low energies with the maximum channel of the pulse heights is about 35, which means the recorded energy of CsI spike events is about 25 keV−95 keV (B. Y.

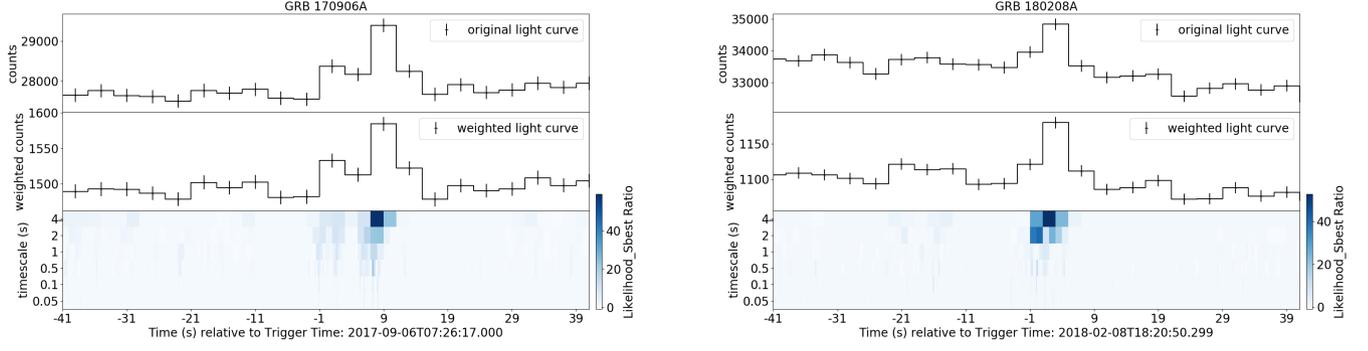

**Figure 6.** Targeted search results for long GRBs: GRB 170906A (left panels) and GRB 180208A (right panels). The raw light curve, weighted light curve and the distribution of the likelihood ratio with different timescales and phases are shown in the top, middle and bottom panels, respectively. The weighted light curve is computed from the raw count rates weighted to maximize the SNR for the best-fit spectrum and location of GRB. T0 is the trigger time given by *Swift* or *Fermi*.

Wu et al. submitted). These spikes in the light curves could have a great impact on the search for bursts, especially for those bursts with soft spectra. As shown in Figure 2, there is a spike around the trigger time of GRB 171216A. Goldstein et al. (2019) have removed the low channels of NaI detectors in the targeted search to avoid false triggers caused by spikes. However, GRBs with soft spectrum are usually most significant in the low energy channels. For example, the events of the BAT GRB 170906C mainly distribute in the low energy range (15 keV−100 keV) (Lien 2017a,b). For these soft GRBs, the method that simply removing the low channels is not appropriate for the targeted search in HE data.

Here we propose a method to eliminate these spike events by evaluating the fraction ($f_{max}$) of the maximum log-Likelihood ratio of a single detector (Max($\mathcal{L}_k$)) to the summed log-Likelihood ratio of all detectors ($\mathcal{L}$). To determine the threshold of this fraction (spike event filter) that we used for this study, the targeted search were applied on the spikes and GRBs of HE CsI data. As shown in Figure 4, the green line and blue line represent the fraction for spikes and GRBs, respectively. Based on this result, we choose the threshold of this fraction ($f_{max}$) to be 0.4, which means that if $f_{max} > 0.4$, we classify it as a spike. Using this spike event filter, about 99% of the spike events are identified and removed while almost all real GRBs are retained.

$$f_{max} = \frac{\text{Max}(\mathcal{L}_k)}{\mathcal{L}}. \tag{8}$$

The log-Likelihood ratio ($\mathcal{L}$) distribution of different energy channels and detectors for the spike events is shown in Figure 3. Data between energy channels 18~30 of the CsI detector #9 is included in the targeted search for GRB 171216A, resulting in a total $\mathcal{L}$ of 47.29 for this event. However, the $\mathcal{L}$ of CsI detector #9 alone is 47.21, with the $f_{max}$ of 0.998, far above the threshold of 0.4. Thus this trigger is classified as spike event, as evident in the light curves (Figure 2).

Using this improved likelihood ratio method, we recovered a short gamma-ray burst, GRB 170906C, through targeted search of HE CsI data. This burst would not trigger if removing the lowest energy channel to avoid spikes. The HE light curves in different energy channels are shown in Figure 5.

Finally, we derived the frequency of occurrence of transients by applying the targeted search over ~1E+06 s (random selection) of HE CsI data, to estimate the false alarm rate (FAR) and the false alarm probability (FAP) for each burst. Similar to GBM, the HE CsI data is very complicated, including non-Gaussian backgrounds, such as instrument noise and many sources from all sky. These phenomena, as well as statistical fluctuations, substantially contribute to the false alarm rate. As mentioned above, some spike events might be misclassified as GRB, which would increase the probability of the occurrence of the false positives. We refined the FAR by removing those time regions with either bad background estimation due to occultation steps from bright persistent sources or quickly-varying counts rate during approach to or exit from the SAA. The false alarm probability is defined as following (see also Kocevski et al. (2018)):

$$P = 1 - e^{-3\lambda_c T}, \tag{9}$$

where $\lambda_c$ and $T$ represent a signal rate by chance and the search time window, respectively.

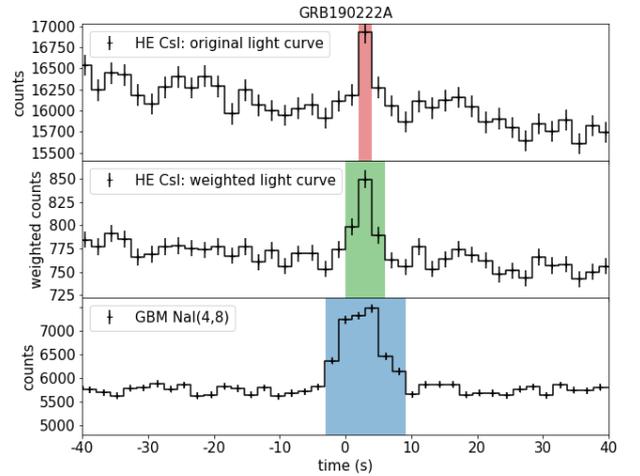

**Figure 7.** GRB 190222A light curves. The HE original light curve (top) and optimized weighted light curve (middle), the GBM light curve (bottom). Color-shaded regions indicate the time intervals where the burst is significant above the background.

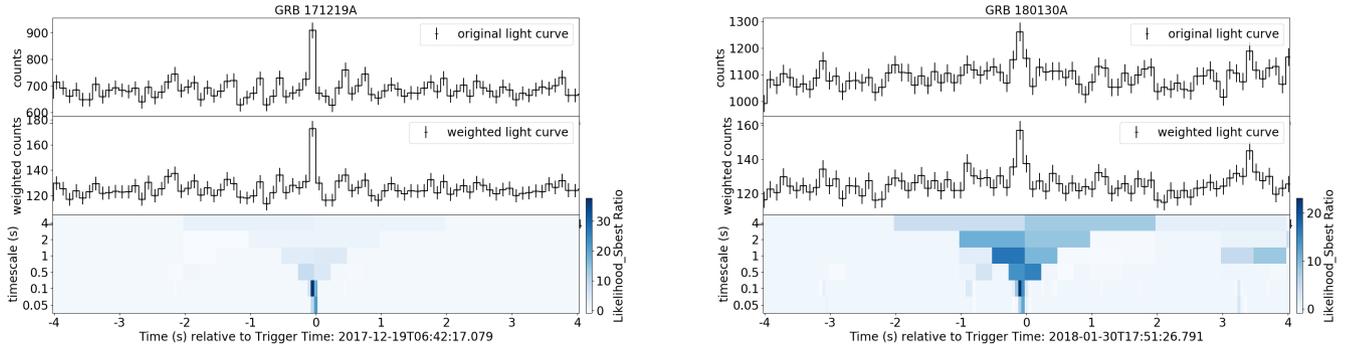

**Figure 8.** Targeted search results for short GRBs: GRB 171219A (left) and GRB 180130A (right). Other captions are same as Figure 6.

### 4.3 Targeted search results

According to our targeted search, 152 of 288 GRBs in three samples are found with $\mathcal{L} > 9$. Out of these 152 GRBs, 112 already triggered the blind search, while the other 40 GRBs are recovered by the targeted search. Among these 40 GRBs, there are 33 long bursts and 8 short bursts. For the long bursts, there are 27 bursts detected by GBM only, 5 bursts detected by BAT only and 1 burst detected by both BAT and GBM. For the short bursts, they were detected by GBM only.

In addition to the targeted search, we also defined the optimized weighted light curve according to the expected counts in each channel and each detector. For channel $i$ and detector $k$, the weight factor is defined as following (see also (Connaughton et al. 2016)):

$$W_{ik} = \frac{S_{ik}}{B_{ik}}, \qquad (10)$$

where $S_{ik}$ and $B_{ik}$ represent the expected counts and background variance in the $i$-th channel and $k$-th detector, respectively.

Then all light curves of all channel and detectors are weighted with $W_{ik}$ and summed up to get the optimized light curve.

The optimized SNR is estimated as:

$$Sig_{\rm opt} = \sum_{i,k=1} \frac{N_{ik} W_{ik}}{\sqrt{B_{ik} W_{ik}^2}}, \qquad (11)$$

where $N_{ik}$ is the background-subtracted data.

We take GRB 170906A and GRB 180208A as examples to show the original light curves (simple summed light curves of all channels and detectors without weight or with the fixed weight of unity), weighted light curves as well as the likelihood ratio maps. As show in Figure 6, the bursts in the weighted light curves are more significant than the original raw light curves. As shown in the lower panels of Figure 6, the likelihood ratio distribution maps contain the likelihood ratios for six timescales and two phase offsets of the targeted search. For each timescale, the likelihood ratio of every data bin is shown according to the best-fit spectral template.

As shown in Figure 7, we compared the GBM and HE light curves for GRB 190222A. The GBM light curve of two triggered detectors (NaI 4, NaI 8) are shown in the bottom panel, where the duration of the burst is marked as the blue-shaded region. However, the original light curve (top panel) and optimized weighted light curve (middle panel) of HE show different duration of this burst. The weighted light curve shows that GRB 190222A is a long GRB, which is consistent with GBM detection. This example demonstrates that the weighted light curve is very useful to unveil light curve components that are too weak to be found in original light curves.

For short GRBs, there are 8 bursts detected by GBM, which are too weak to be found in blind search but rediscovered through the targeted search analysis in the HE data. Similar plots for two bursts, GRB 171219A and GRB 180130A, are shown in Figure 8, which show that these two bursts were recovered only in shorter timescales. For GRB 180130A, it is difficult to identify the burst in the original light curve (top panel), but it is apparent in the weighted light curve with maximized SNR (middle panel).

The remaining 136 bursts in the sample can not trigger the detection threshold in the targeted search. We did a Monte Carlo simulation for GRB 180102A, which was selected from the remaining burst samples, with the location from BAT (Lien et al. 2016) and the spectrum from GBM (with a wider energy range (Gruber et al. 2014; von Kienlin et al. 2014; Bhat et al. 2016)). The location of GRB 180102A is shown in Figure 9 (left panel), which is in the field of view of HE CsI gamma-ray monitoring. Figure 9 shows the expected counts rate of each CsI detector for GRB 180102A on right panel. Some CsI detectors are expected to deposit more counts than others, especially the detector #17. However, the total expected count (∼367) is well within the 2-$\sigma$ statistical fluctuation of the background (∼32723) in 4 s duration (the maximum timescale in our search). Therefore, there is no data bin exceeding the background variation significantly for both blind search and targeted search.

For each burst detected by BAT or GBM during the first two years of HXMT, which triggered HE in blind search, we also applied the targeted search to identify coincident signals in HE CsI detectors.

For all bursts in our samples, the relation between the likelihood ratios and the optimized SNR is shown in Figure 10, in which the log of likelihood ratio is proportional to the optimized SNR. Those bursts which triggered HE in both blind search and targeted search (green triangles) and which were only recovered in targeted search (blue circles), are differentiated by their likelihood ratios. Generally, when the likelihood ratios are greater than ∼20 (equivalent to SNR of ∼6), most bursts could successfully trigger the blind search.

Their fluence versus $T_{90}$ in the 10−1000 keV energy band of the GBM samples is shown in Figure 12. There are 188 bursts of GBM (left panel), 70 of which are found in neither blind nor targeted search of HE CsI data. Among HE-detected 118 GRBs (right panel), 34 are only triggered by the targeted search. The peak counts rate distribution of HE GRBs is shown in Figure 13. The results of the targeted search for the entire samples are summarized in Tabel 4 and Tabel 3.

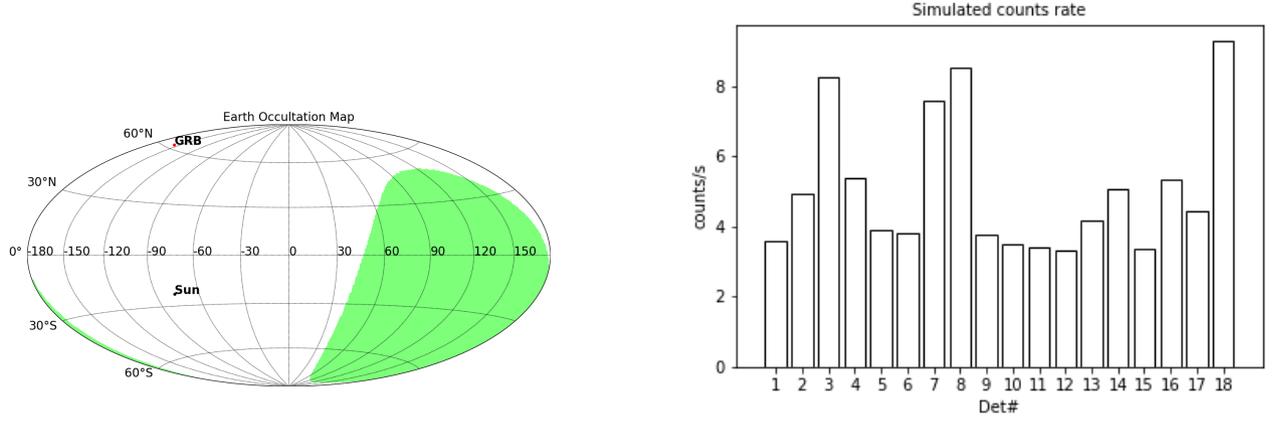

**Figure 9.** *Insight*-HXMT observation of GRB 180102A. (Left panel) Field of View of *Insight*-HXMT/HE gamma-ray monitoring. The sky region shielded by the Earth at the trigger time is shown in green. The red point and black point represent the location of GRB 180102A and the Sun respectively. (Right panel) The expected counts rate of each HE CsI detector for GRB 180102A is calculated using the GRB location given by BAT, the spectrum by GBM (Gruber et al. 2014; von Kienlin et al. 2014; Bhat et al. 2016) and the detector response of HE. The total expected counts of HE are ∼1183 in the 13 s duration of GRB 180102A (right).

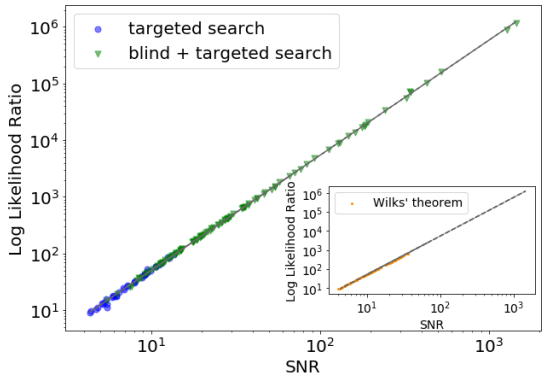

**Figure 10.** The relation between likelihood ratios and optimized SNR of 152 GRBs in our samples. The green triangles represent the bursts that can be found in both targeted search and blind search, and the blue points are the bursts that can only be detected in targeted search. The black dashed line is the fitting results of green and blue data points. In the inset panel, the black dashed line also represents the fitting results of green and blue data points while the orange points are calculated based on Wilks' theorem. We note that both the black dash line and the trend of orange points tightly approximate to the line of equivalence between the likelihood ratio and the half of the square of the optimized SNR.

## 5 SEARCH FOR GW EM

GW triggers from the LIGO–Virgo Gravitational wave Transient Catalog of compact binary coalescences observed in the second observing run (O2) (Abbott et al. 2019) and the first half of the third observing run (O3a) (Abbott et al. 2020) are listed in Table 5, which pass the FAR threshold of 1 per 30 days (12.2 per year) and 2 per one year during O2 and O3a, respectively. The maximum astrophysical origin probabilities in the three pipelines (GstLAL, PyCBC and PyCBC BBH) of all events are also shown in Table 5.

The HE coverage of the LIGO/VIRGO GW candidates is estimated through public HEALPix (Gorski et al. 2005) sky location maps from GWTC-1 (LIGO Scientific & Virgo Collaboration 2018) and GWTC-2 (LIGO Scientific & Virgo Collaboration 2020). For O3a, we make use of the software package ligo.skymap [1] to convert the HEALPix FITS file from multi-resolution UNIQ indexing to the more common IMPLICIT indexing. The flattened fits are used to generate skymaps which had corresponding HE CsI data. *Insight*-HXMT were almost fully occulted by the Earth for two GW candidates with the location region coverage of only 0.09% and 4.38%. *Insight*-HXMT were turned off due to the SAA passage for these GW events: GW 190421_213856, GW 190424_180648, GW 190720_000836, GW 190727_060333, GW 190828_063405, GW 190924_021846. For the remaining GW events in the sample of this paper, the location region probability covered by HE ranges from 15% to 100%. As an example, the HE coverage sky map of the GW190521_074359 is shown in Figure 14.

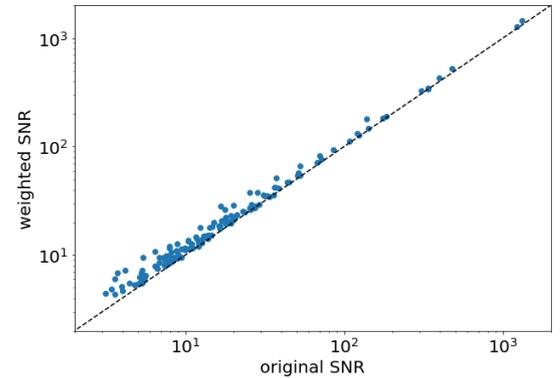

**Figure 11.** The optimized weighted SNR versus original SNR in our samples. The black dashed line represents the equivalence between weighted SNR and original SNR.

We examined the time offsets between the GW triggers from the catalog and GRBs from the blind search of HE data and found no GRBs associated with GW events. Thus we further used the targeted search to search for counterparts candidates of GW triggers. The targeted search was performed on HE EVT data for any burst candidates on six time scales of 0.05 s, 0.1 s, 0.5 s, 1 s, 2 s, 4 s within ±30 s of the GW triggers time, including GRB 170817A. Once a candidate is found, the optimized and weighted light curves will be produced by the coherent search pipeline and the nominal SNR will be calculated accordingly. However, as described in the previous section, the real false alarm rate (FAR) will be estimated by applying the targeted search to the background data around the GW triggers with the same search parameters and criteria (including the log-likelihood ratio trigger threshold of 9), in order to estimate the real significance of the potential counterpart. In this work, no significant counterpart has been found. We calculated the expected counts of GRB 170817A with the latest version of the detector response (Luo et al. 2020), resulting in only ∼ 30 counts in HE detectors, which is consistent with the non-detection reported in Li et al. (2018).

---

[1] https://lscsoft.docs.ligo.org/ligo.skymap/

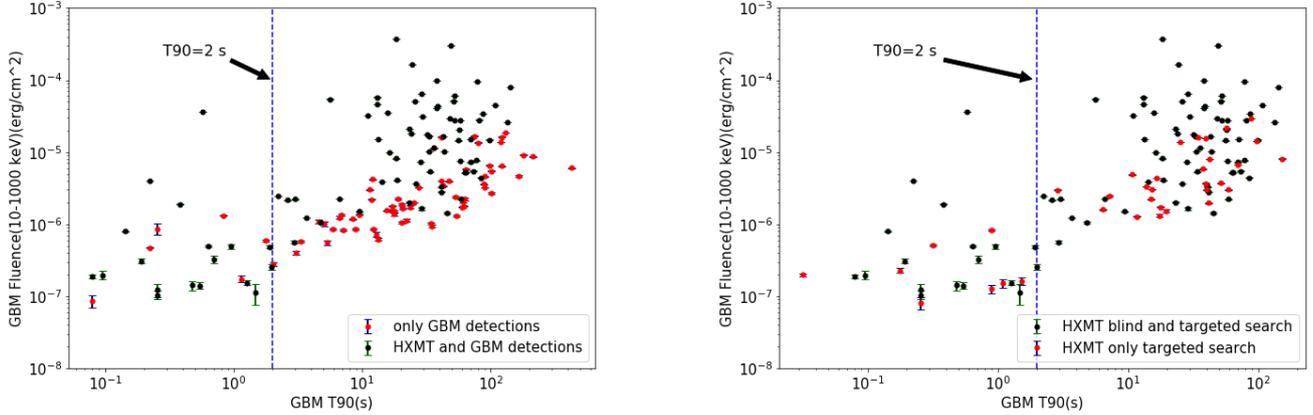

**Figure 12.** The fluence versus $T_{90}$ of GBM GRB samples. The black points represent the bursts that were detected by both GBM and HE, while the red points represent the bursts that only triggered GBM (left). The black points represent the bursts that were detected in both HE blind and target search, while the bursts only detected in HE targeted search are represented by red points (right). The error bars represent error of the fluence.

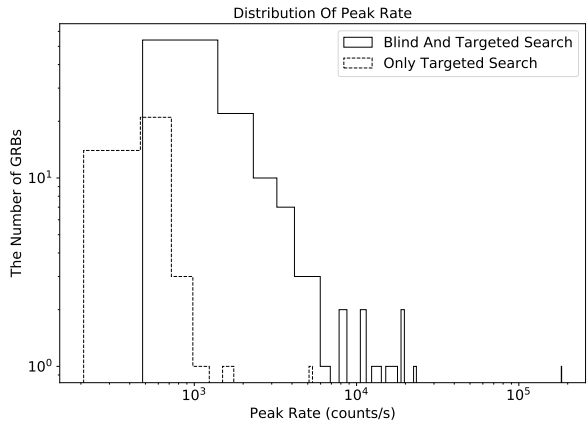

**Figure 13.** The peak counts rate distribution of GRB samples. The solid black line represents the bursts triggered the blind search and targeted search with HE data; the dotted line represents the bursts which are only triggered the targeted search.

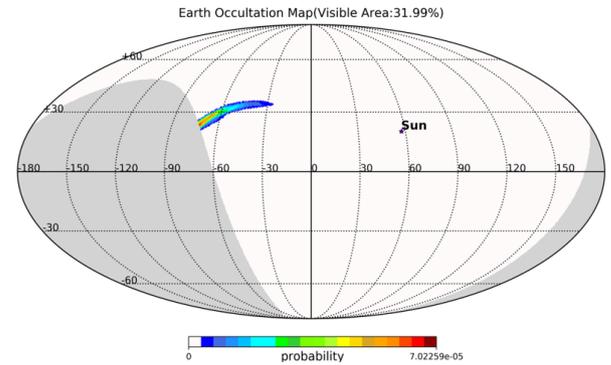

**Figure 14.** The HE coverage sky map of LIGO/VIRGO candidate GW 190521_074359, where the gray shadow indicates the area blocked by the Earth. The visible area is about 32 percent of the total GW localization probability and the maximum probability location is ra=287.40°, decc=23.40°. The GW location probability is represented as color scale.

The 5-sigma upper limit flux in 0.2–3 MeV (incident energy) was estimated using the original light curve and optimized weighted light curve as following: Firstly, assuming the GW counterpart GRB with three typical Band spectral models, one typical duration time scale (1 s), and one location from the center of the LIGO-Virgo location probability map, we can compute the expected counts in each channel and each detector (the initial value of normalization parameter of spectrum is set to be 1). To derive the optimized weighted light curve, the weighted factor can be estimated with the expected counts and background of GW trigger time. As the optimized SNR varies with the normalization of the spectrum, we calculate the corresponding normalization when optimized SNR equals 5-sigma. Using this normalization together with other spectral parameters, we can finally compute the 5-sigma upper limit flux in 0.2–3 MeV.

Taking GW190412 as an example, we estimated the upper limit with soft Band spectrum and time scale of 1 s to be 7.15E-07 erg cm$^{-2}$ s$^{-1}$ using the original light curve, whereas the upper limit can reach 5.69E-07 erg cm$^{-2}$ s$^{-1}$ with the optimized weighted light curve. The optimized upper limit flux with three typical GRB Band spectral models of all GW triggers are listed in Table 5.

## 6 DISCUSSION

The results outlined in section 4.3 demonstrate that the targeted search is an effective and reliable method to identify weak transient signals, including both LGRBs and SGRBs. The targeted search increases sensitivity by coherently accumulating HE multiple-detector multiple-channel data based on the expected model rates considering incident angle, detector response and GRB spectrum. We find a total of 40 bursts (above the likelihood ratio threshold of 9) from June 15, 2017 to June 30, 2019, which did not trigger the blind search of HXMT CsI data. The trigger time of those 40 bursts is consistent with that of GBM or BAT, which are shown in Figures 6 and 8.

The optimized light curves weighted according to the model rates with the highest likelihood ratio can substantially enhance the significance of the burst and uncover the weak components of burst, as shown in Figure 7.

There are some GRBs in our samples, like GRB 180102A as shown in Figure 9, which fall in the field of view of HE, but still did not trigger the targeted search (left panel). The distribution of very little expected counts of 18 CsI detectors for GRB180102A on the right panel of Figure 9, showing that there is almost no counts above 100 keV, suggests that this burst has a soft spectrum, which is consistent with the measurements of GBM and BAT.

The likelihood ratio and optimized SNR for our GRB sample are shown in Figure 10. There are no obvious boundaries to separate the BAT or GBM detected bursts that also triggered HE in blind search and those which were only recovered in targeted search. However, it appears that only targeted search can find the bursts with SNR ≤ 6. We find that, as shown in Figure 11, the relation between optimized SNR and log likelihood ratio is quite consistent with Wilks' theorem, and the weighted optimized SNR is almost always larger than the original SNR (i.e. weight of each detector and each channel is set to be 1).

The energy fluence of GBM samples versus their $T_{90}$, as shown in Figure 12, demonstrates that for SGRBs, the sensitivity of HE and GBM is very close to each other, while for LGRBs, GBM is more sensitive than HE (left). The right panel shows that the minimum fluence could reach 8.14E-08 erg/cm$^2$ (10 keV−1 MeV) in duration of 0.256 s in targeted search, which suggests that this method could complement blind search effectively.

The peak counts rate distribution of detected GRB samples, shown in Figure 13 indicates that the targeted search method is more sensitive to detect weak bursts. The minimum peak rate of blind search is 480 counts/s, and the targeted search method can reach 238 counts/s.

Targeted search results are summarized in Table 3 and 4. The proportion of the detected GRBs of HE in GBM samples is higher than in BAT samples, which shows BAT is more sensitive to detect much weaker GRBs in 15−150 keV. The proportion of the detected SGRBs (in all SGRBs) is higher than LGRBs (in all LGRBs), due to the detection energy range of HE (i.e. 200 to 3000 keV), which is more sensitive to GRBs with harder spectra.

## 7 SUMMARY

In this paper, we report the search results for GRB and GW EM in *Insight*-HXMT/HE data. Both blind search and targeted search have been implemented in our data processing pipeline.

The targeted coherent search method is applied to the EVT data of HE CsI detectors, which can recover true astrophysical signals that are too weak to trigger the blind search. This coherent method was originally developed by (Blackburn et al. 2013), improved by (Goldstein et al. 2016, 2019) and examined by (Kocevski et al. 2018). In this paper, we proposed a method to filter out false triggers caused by spikes in the light curves, which could be used in other GRB detectors, such as *Fermi*/GBM and POLAR, which suffers from similar spikes.

We find that the targeted search method can successfully recover some GRBs which are too weak to trigger the blind search. We show the sky area which monitored by HE at the trigger time and the expected counts rate of GRB 180102A (detected by GBM and BAT), which was not recovered by the targeted search of HE data. We studied the relation between the likelihood ratios and the optimized SNR for our GRB sample and found that it is not equal to but quite consistent with the Wilks' theorem. Our work also demonstrate that the targeted search method is crucial to increasing the HE CsI sensitivity to gamma-ray transients. We also find minimum energy fluence of the HE CsI targeted search could reach 8.14E-08 erg/cm$^2$ (10 keV−1 MeV) in the time scale of 0.256 s. Additionally, we calculate the GRB detection sensitivity of *Insight*-HXMT/HE over the CsI observing range (0.2−3 MeV), which is about 1.5E-08 erg/cm$^2$. Finally, our targeted search pipeline is also performed on HE CsI data for seeking EM counterparts of GW events including the first neutron star binary merger event (GW170817). Our results show that the targeted search (optimized light curve) can achieve a higher sensitivity for gamma-ray bursts and thus provide more stringent constraint on the burst emission associated with GW events and FRBs.

## ACKNOWLEDGEMENTS

We thank the anonymous reviewer for very helpful comments and suggestions. This work made use of the data from the *Insight*-HXMT mission, a project funded by China National Space Administration (CNSA) and the Chinese Academy of Sciences (CAS). The authors thank supports from the National Program on Key Research and Development Project (Grant No. 2016YFA0400801), the National Natural Science Foundation of China under Grants No. U1838113, U1938102, 11503029, U1838201, U1838202 and the Strategic Priority Research Program on Space Science of the Chinese Academy of Sciences (Grant No. XDB23040400).

## DATA AVAILABILITY

The data underlying this article will be shared on reasonable request to the corresponding author.

## REFERENCES

Abbott B. P., et al., 2017a, Phys. Rev. Lett., 119, 161101
Abbott B. P., et al., 2017b, Astrophys. J., 848, L12
Abbott B. P., et al., 2019, Phys. Rev. X, 9, 031040
Abbott R., et al., 2020, arXiv : 2010.14527
Band D., et al., 1993, ApJ., 413, 281
Barthelmy S. D., et al., 2005, Space Sci. Rev., 120, 143
Bhat P. N., et al., 2016, Astrophys. J. Suppl., 223, 28
Blackburn L., Briggs M. S., Camp J., Christensen N., Connaughton V., Jenke P., Veitch J., 2013, in Proceedings, 4th International Fermi Symposium: Monterey, California, USA, October 28-November 2, 2012. (arXiv:1303.2174)
Blackburn L., Briggs M. S., Camp J., Christensen N., Connaughton V., Jenke P., Remillard R. A., Veitch J., 2015, Astrophys. J. Suppl., 217, 8
Burns E., et al., 2019, Astrophys. J., 871, 90
Cao X., et al., 2019, Sci. China Phys. Mech. Astron. 63, 249504 (2020)
Chen Y., et al., 2019, Sci. China Phys. Mech. Astron. 63, 249505 (2020)

**Table 3.** GRBs found by the targeted search in *Insight*-HXMT/HE data.

| UTC Time | Instrument | $\Delta t^a$ (s) | Timescale (s) | Spectral Template | Detector Gain Mode | $\mathcal{L}$ | FAR (Hz) | FAP |
|---|---|---|---|---|---|---|---|---|
| 2017-07-11T22:20:25.000 | BAT+GBM | 1.00 | 4.00 | hard Band | normal gain | 13.24 | 2.20E-05 | 6.59E-03 |
| 2017-07-18T03:39:30.168 | GBM | -1.00 | 4.00 | norm Band | normal gain | 24.57 | 9.90E-06 | 2.97E-03 |
| 2017-08-31T04:18:11.132 | GBM | 26.50 | 1.00 | soft Band | normal gain | 85.82 | 2.78E-06 | 8.33E-04 |
| 2017-09-06T07:26:17.000 | BAT | 7.00 | 4.00 | soft Band | normal gain | 59.22 | 8.92E-06 | 2.67E-03 |
| 2017-09-26T12:39:41.225 | GBM | -1.00 | 2.00 | soft Band | normal gain | 51.6 | 4.63E-06 | 1.39E-03 |
| 2017-10-13T08:24:42.108 | GBM | 7.00 | 4.00 | norm Band | normal gain | 67.25 | 5.95E-06 | 1.78E-03 |
| 2017-10-29T00:28:49.091 | GBM | -0.50 | 1.00 | soft Band | normal gain | 19.02 | 5.73E-05 | 1.70E-02 |
| 2017-11-06T11:56:49.272 | GBM | 2.00 | 1.00 | norm Band | low gain | 9.02 | 2.93E-04 | 8.40E-02 |
| 2017-12-07T01:18:42.452 | GBM | 0.00 | 0.50 | hard Band | normal gain | 18.24 | 4.49E-05 | 1.35E-03 |
| 2017-12-19T06:42:17.079 | GBM | -0.05 | 0.10 | hard Band | normal gain | 37.75 | 5.62E-05 | 1.68E-03 |
| 2018-01-30T17:51:26.791 | GBM | -0.10 | 0.10 | norm Band | normal gain | 23.12 | 2.94E-04 | 8.77E-03 |
| 2018-02-08T18:20:50.299 | GBM | 1.00 | 4.00 | soft Band | normal gain | 52.73 | 9.26E-06 | 2.78E-03 |
| 2018-02-22T13:10:03.000 | BAT | 3.00 | 4.00 | norm Band | normal gain | 42.12 | 1.35E-05 | 4.04E-03 |
| 2018-03-06T23:20:33.933 | GBM | 1.00 | 4.00 | hard Band | normal gain | 21.17 | 1.01E-05 | 3.02E-03 |
| 2018-03-09T07:43:10.615 | GBM | 7.00 | 4.00 | norm Band | normal gain | 46.37 | 1.35E-05 | 4.04E-03 |
| 2018-04-20T00:45:09.919 | GBM | -1.00 | 4.00 | hard Band | normal gain | 33.18 | 8.54E-06 | 2.56E-03 |
| 2018-05-04T03:15:53.830 | GBM | 5.00 | 4.00 | norm Band | normal gain | 44.38 | 1.35E-05 | 4.04E-03 |
| 2018-05-17T07:24:21.380 | GBM | 0.00 | 0.50 | hard Band | normal gain | 10.51 | 3.00E-05 | 8.96E-03 |
| 2018-06-20T08:34:58.000 | BAT | 9.00 | 1.00 | soft Band | normal gain | 13.02 | 1.20E-04 | 3.54E-02 |
| 2018-07-18T04:49:00.200 | GBM | -1.00 | 4.00 | norm Band | normal gain | 42.8 | 1.35E-05 | 4.04E-03 |
| 2018-08-01T06:37:03.514 | GBM | 0.25 | 0.50 | norm Band | normal gain | 17.15 | 1.07E-04 | 3.21E-03 |
| 2018-08-10T06:40:46.744 | GBM | -1.00 | 4.00 | norm Band | low gain | 11.24 | 3.00E-04 | 8.61E-02 |
| 2018-08-16T02:07:18.913 | GBM | 19.00 | 2.00 | soft Band | normal gain | 96.22 | 2.78E-06 | 8.33E-04 |
| 2018-08-22T13:28:34.050 | GBM | 1.50 | 1.00 | hard Band | normal gain | 31.14 | 4.50E-06 | 1.35E-03 |
| 2018-09-22T11:03:52.023 | GBM | 1.00 | 4.00 | norm Band | normal gain | 58.31 | 1.12E-05 | 3.36E-03 |
| 2018-09-25T09:46:30.545 | GBM | -0.35 | 0.10 | hard Band | normal gain | 12.66 | 1.61E-04 | 4.82E-03 |
| 2018-11-17T15:55:18.057 | GBM | 9.00 | 4.00 | norm Band | normal gain | 84.59 | 2.97E-06 | 8.92E-04 |
| 2018-12-17T15:56:59.038 | GBM | 1.00 | 4.00 | norm Band | normal gain | 18.38 | 3.00E-05 | 8.96E-03 |
| 2018-12-24T23:41:47.453 | GBM | 0.05 | 0.10 | norm Band | normal gain | 17.49 | 3.05E-04 | 9.11E-03 |
| 2019-02-22T07:29:35.526 | GBM | 3.00 | 2.00 | norm Band | normal gain | 40.03 | 4.50E-06 | 1.35E-03 |
| 2019-03-04T19:37:23.342 | GBM | 0.50 | 1.00 | hard Band | normal gain | 13.55 | 1.15E-05 | 3.43E-03 |
| 2019-04-07T16:07:26.493 | GBM | 0.00 | 1.00 | norm Band | normal gain | 17.43 | 4.74E-05 | 1.41E-02 |
| 2019-04-28T18:48:12.460 | GBM | 5.00 | 4.00 | norm Band | low gain | 26.57 | 1.87E-05 | 5.61E-03 |
| 2019-05-05T01:14:09.330 | GBM | -0.03 | 0.05 | hard Band | low gain | 11.14 | 2.08E-04 | 6.22E-03 |
| 2019-05-08T19:22:50.400 | GBM | 1.00 | 4.00 | norm Band | normal gain | 27.1 | 1.56E-05 | 4.68E-03 |
| 2019-05-10T02:52:13.232 | GBM | 1.00 | 4.00 | norm Band | normal gain | 49.82 | 4.50E-06 | 1.35E-03 |
| 2019-05-25T00:45:47.652 | GBM | 0.50 | 1.00 | hard Band | normal gain | 30.7 | 4.49E-05 | 1.35E-03 |
| 2019-06-04T14:57:15.000 | BAT | 1.00 | 4.00 | hard Band | normal gain | 63.94 | 2.97E-06 | 8.92E-04 |
| 2019-06-09T07:34:05.259 | GBM | 17.00 | 4.00 | hard Band | normal gain | 42.99 | 4.50E-06 | 1.35E-03 |
| 2019-06-30T23:52:59.000 | BAT | 1.00 | 4.00 | soft Band | normal gain | 48.73 | 1.35E-05 | 4.04E-03 |

$^a$ The offset between the BAT/GBM trigger time (T0) and the center of the detection window of the highest $\mathcal{L}$ for the bursts.

**Table 4.** Summary of results for the GRB samples

| Samples | visible | | | | | | invisible | | proportion | | | |
|---|---|---|---|---|---|---|---|---|---|---|---|---|
| | detected | | | | undetected | | SAA | Earth occlusion | detected | | SAA | Earth occlusion |
| | long GRBs | | short GRBs | | long GRBs | short GRBs | – | – | long GRBs | short GRBs | – | – |
| | T* | BT** | T | BT | – | – | – | – | – | – | – | – |
| GBM | 26 | 66 | 8 | 18 | 63 | 7 | 89 | 155 | 0.60 | 0.78 | 0.2 | 0.35 |
| BAT | 5 | 14 | 0 | 3 | 48 | 3 | 19 | 53 | 0.28 | 0.5 | 0.13 | 0.36 |
| GBM BAT | 1 | 9 | 0 | 2 | 14 | 1 | 9 | 14 | 0.45 | 0.67 | 0.18 | 0.28 |

* The bursts number of targeted search
** The bursts number of blind search and targeted search

**Table 5.** *Insight*-HXMT/HE observation of GW events in GWTC-1 and GWTC-2 (Abbott et al. 2019, 2020).

| GW Event | Time (UTC) | $P_{astro}$[a] | Ra[b] (deg) | Dec[c] (deg) | HE Coverage[d] | Upper Limit[e] (erg cm$^{-2}$ s$^{-1}$) | Upper Limit[f] (erg cm$^{-2}$ s$^{-1}$) | Upper Limit[g] (erg cm$^{-2}$ s$^{-1}$) |
|---|---|---|---|---|---|---|---|---|
| GW170729 | 2017-07-29T18:56:29 | 0.98 | 332.67 | -69.89 | 94.49 % | 6.09E-07 | 4.47E-07 | 3.19E-07 |
| GW170809 | 2017-08-09T08:28:21 | 1 | 14.15 | -29.48 | 100 % | 6.04E-07 | 3.39E-07 | 2.31E-07 |
| GW170814 | 2017-08-14T10:30:43 | 1 | 47.32 | -44.45 | 99.37 % | 5.90E-07 | 2.86E-07 | 1.90E-07 |
| GW170817 | 2017-08-17T12:41:04 | 1 | 196.3 | -21.3 | 100 % | 5.45E-07 | 5.00E-07 | 3.75E-07 |
| GW170818 | 2017-08-18T02:25:09 | 1 | 341.28 | 22.39 | 100 % | 2.91E-07 | 9.69E-08 | 5.93E-08 |
| GW170823 | 2017-08-23T13:13:58 | 1 | 58.39 | 42.61 | 42.17 % | 2.66E-07 | 8.86E-08 | 5.37E-08 |
| GW190408_181802 | 2019-04-08T18:18:02 | 1 | 351.02 | 53.91 | 99.52 % | 3.56E-07 | 1.27E-07 | 7.85E-08 |
| GW190412 | 2019-04-12T05:30:44 | 1 | 212.48 | 32.18 | 4.38 % | 6.35E-07 | 7.00E-07 | 5.69E-07 |
| GW190413_052954 | 2019-04-13T05:29:54 | 0.98 | 66.84 | -44.99 | 79.74 % | 3.33E-07 | 1.13E-07 | 6.96E-08 |
| GW190413_134308 | 2019-04-13T13:43:08 | 0.98 | 156.36 | -32.44 | 96.82 % | 6.23E-07 | 4.81E-07 | 3.44E-07 |
| GW190421_213856 | 2019-04-21T21:38:56 | 1 | 205.3 | -46.38 | - | - | - | - |
| GW190424_180648 | 2019-04-24T18:06:48 | 0.91 | 225 | -40.33 | - | - | - | - |
| GW190425 | 2019-04-25T08:18:05 | - | 71.72 | -24.79 | 58.15 % | 3.10E-07 | 9.87E-08 | 5.96E-08 |
| GW190426_152155 | 2019-04-26T15:21:55 | - | 352.57 | 85.29 | 98.42 % | 5.02E-07 | 2.18E-07 | 1.42E-07 |
| GW190503_185404 | 2019-05-03T18:54:04 | 1 | 137.81 | -41.06 | 0.09 % | 7.67E-07 | 7.05E-07 | 5.30E-07 |
| GW190512_180714 | 2019-05-12T18:07:14 | 1 | 250.18 | -26.61 | 99.71 % | 7.97E-07 | 7.04E-07 | 5.17E-07 |
| GW190513_205428 | 2019-05-13T20:54:28 | 1 | 286.7 | -30.43 | 16.25 % | 5.26E-07 | 5.51E-07 | 4.36E-07 |
| GW190514_065416 | 2019-05-14T06:54:16 | 0.96 | 357.91 | 76.17 | 28.32 % | 6.48E-07 | 2.73E-07 | 1.77E-07 |
| GW190517_055101 | 2019-05-17T05:51:01 | 1 | 229.11 | -46.47 | 99.97 % | 7.34E-07 | 4.44E-07 | 3.05E-07 |
| GW190519_153544 | 2019-05-19T15:35:44 | 1 | 353.39 | 43.31 | 67.58 % | 9.24E-07 | 6.21E-07 | 4.32E-07 |
| GW190521 | 2019-05-21T03:02:29 | 1 | 5.06 | -66.26 | 98.15 % | 6.53E-07 | 3.45E-07 | 2.32E-07 |
| GW190521_074359 | 2019-05-21T07:43:59 | 1 | 287.4 | 23.4 | 31.99 % | 5.08E-07 | 5.85E-07 | 5.06E-07 |
| GW190527_092055 | 2019-05-27T09:20:55 | 0.99 | 23.76 | -68.59 | 93.89 % | 3.06E-07 | 1.03E-07 | 6.33E-08 |
| GW190602_175927 | 2019-06-02T17:59:27 | 1 | 80.07 | -41.01 | 58.62 % | 3.55E-07 | 1.06E-07 | 6.36E-08 |
| GW190620_030421 | 2019-06-20T03:04:21 | 1 | 255.06 | 23.72 | 96.44 % | 6.08E-07 | 2.96E-07 | 1.97E-07 |
| GW190630_185205 | 2019-06-30T18:52:05 | 1 | 341.02 | -7.86 | 93.56 % | 9.42E-07 | 6.25E-07 | 4.36E-07 |
| GW190701_203306 | 2019-07-01T20:33:06 | 1 | 37.79 | -7.29 | 100 % | 5.57E-07 | 2.57E-07 | 1.69E-07 |
| GW190706_222641 | 2019-07-06T22:26:41 | 1 | 145.37 | 22.75 | 92.05 % | 6.71E-07 | 3.54E-07 | 2.38E-07 |
| GW190707_093326 | 2019-07-07T09:33:26 | 1 | 148.89 | -30.69 | 74.96 % | 8.54E-07 | 4.67E-07 | 3.16E-07 |
| GW190708_232457 | 2019-07-08T23:24:57 | 1 | 201.58 | 65.23 | 79.03 % | 4.10E-07 | 1.77E-07 | 1.15E-07 |
| GW190719_215514 | 2019-07-19T21:55:14 | 0.82 | 155.04 | 41.31 | 77.56 % | 2.81E-07 | 9.99E-08 | 6.20E-08 |
| GW190720_000836 | 2019-07-20T00:08:36 | 1 | 297.55 | 36.05 | - | - | - | - |
| GW190727_060333 | 2019-07-27T06:03:33 | 1 | 98.79 | -39.16 | - | - | - | - |
| GW190728_064510 | 2019-07-28T06:45:10 | 1 | 312.32 | 7.56 | 33.23 % | 6.78E-07 | 6.11E-07 | 4.57E-07 |
| GW190731_140936 | 2019-07-31T14:09:36 | 0.97 | 190.01 | -58.82 | 94.99 % | 6.30E-07 | 3.81E-07 | 2.61E-07 |
| GW190803_022701 | 2019-08-03T02:27:01 | 0.99 | 93.6 | -7.78 | 100 % | 3.50E-07 | 1.19E-07 | 7.17E-08 |
| GW190814 | 2019-08-14T21:10:39 | 1 | 12.7 | -24.87 | 100 % | 6.81E-07 | 3.72E-07 | 2.52E-07 |
| GW190828_063405 | 2019-08-28T06:34:05 | 1 | 140.98 | -23.77 | - | - | - | - |
| GW190828_065509 | 2019-08-28T06:55:09 | 1 | 145.47 | -49.51 | 90.47 % | 3.80E-07 | 1.47E-07 | 9.31E-08 |
| GW190909_114149 | 2019-09-09T11:41:49 | 0.89 | 43.07 | 40.72 | 63.46 % | 5.74E-07 | 2.01E-07 | 1.25E-07 |
| GW190910_112807 | 2019-09-10T11:28:07 | 1 | 113.2 | 5.9 | 85.53 % | 6.61E-07 | 2.06E-07 | 1.22E-07 |
| GW190915_235702 | 2019-09-15T23:57:02 | 1 | 195.47 | 37.26 | 100 % | 6.94E-07 | 3.68E-07 | 2.49E-07 |
| GW190924_021846 | 2019-09-24T02:18:46 | 1 | 162.51 | -21.22 | - | - | - | - |
| GW190929_012149 | 2019-09-29T01:21:49 | 1 | 101.6 | -34.32 | 85.54 % | 6.51E-07 | 4.93E-07 | 3.52E-07 |
| GW190930_133541 | 2019-09-30T13:35:41 | 0.99 | 318.87 | 40.82 | 62.15 % | 6.44E-07 | 3.95E-07 | 2.71E-07 |

[a] The $P_{astro}$ values are the maximum reported values, obtained from the GstLAL, PyCBC or PyCBC BBH pipelines.
[b] The maximum probability of Right Ascension (ra).
[c] The maximum probability of Declination (dec).
[d] There are no specific coverage and upper-limit values because HE passaged through SAA at the triggers time.
[e] The 5-sigma upper limit flux in 0.2−3 MeV (incident energy) with hard Band spectrum.
[f] The 5-sigma upper limit flux in 0.2−3 MeV (incident energy) with normal Band spectrum.
[g] The 5-sigma upper limit flux in 0.2−3 MeV (incident energy) with soft Band spectrum.

Connaughton V., et al., 2015, Astrophys. J. Suppl., 216, 32
Connaughton V., et al., 2016, The Astrophysical Journal, 826, L6
Fishman G., Austin R., 1977, Nuclear Instruments and Methods, 140, 193
Goldstein A., Burns E., Hamburg R., Connaughton V., Veres P., Briggs M. S., Hui C. M., 2016, arXiv:1612.02395.
Goldstein A., et al., 2017, Astrophys. J., 848, L14
Goldstein A., et al., 2019, arXiv:1903.12597
Gorski K. M., Hivon E., Banday A. J., Wandelt B. D., Hansen F. K., Reinecke M., Bartelman M., 2005, Astrophys. J., 622, 759
Gruber D., et al., 2014, Astrophys. J. Suppl., 211, 12
Guidorzi C., et al., 2020a, Astron. Astrophys., 637, A69
Guidorzi C., et al., 2020b, Astron. Astrophys., 642, A160
Kocevski D., et al., 2018, Astrophys. J., 862, 152
Kouveliotou C., et al., 1992, Astrophysical Journal, 392, 179

LIGO Scientific & Virgo Collaboration 2018, Sky localization probability maps (skymaps) release for GWTC-1, https://dcc.ligo.org/LIGO-P1800381/public, https://dcc.ligo.org/LIGO-P1800381/public

LIGO Scientific & Virgo Collaboration 2020, Sky localization probability maps (skymaps) release for GWTC-2, https://dcc.ligo.org/LIGO-P2000223/public/, https://dcc.ligo.org/LIGO-P2000223/public/

Li T., et al., 2018, Sci. China Phys. Mech. Astron., 61, 031011

Li X., et al., 2019, Ground-based calibration and characterization of the HE detectors for Insight-HXMT (arXiv:1910.04359)

Li C. K., et al., 2021, Nature Astronomy, 5, 378–384

Lien A. Y., 2017a, GCN CIRCULAR 21832, https://gcn.gsfc.nasa.gov/gcn3/21832.gcn3, https://gcn.gsfc.nasa.gov/gcn3/21832.gcn3

Lien A. Y., 2017b, GRB 170906C, https://gcn.gsfc.nasa.gov/notices_s/770981/BA/, https://gcn.gsfc.nasa.gov/notices_s/770981/BA/

Lien A., et al., 2016, Astrophys. J., 829, 7

Liu C. Z., et al., 2019, Sci. China Phys. Mech. Astron. 63, 249503 (2020)

Luo Q., et al., 2020, Journal of High Energy Astrophysics, 27, 1

Meegan C., et al., 2009, The Astrophysical Journal, 702, 791

Savchenko V., et al., 2017, Astrophys. J., 848, L15

Xiao S., et al., 2020, Journal of High Energy Astrophysics, 26, 58

Zhang S. N., 2009, in Bulletin of the American Astronomical Society. p. 474

Zhang S., Lu F. J., Zhang S. N., Li T. P., 2014, in Space Telescopes and Instrumentation 2014: Ultraviolet to Gamma Ray. SPIE, pp 588 – 595, doi:10.1117/12.2054144, https://doi.org/10.1117/12.2054144

Zhang S. N., et al., 2019, Sci. China Phys. Mech. Astron. 63, 249502 (2020)

von Kienlin A., et al., 2014, Astrophys. J. Suppl., 211, 13

von Kienlin A., et al., 2020, The Astrophysical Journal, 893, 46

This paper has been typeset from a T<sub>E</sub>X/L<sup>A</sup>T<sub>E</sub>X file prepared by the author.